\documentclass[a4paper,11pt]{article}

\usepackage{here}
\usepackage{subfig}
\usepackage[dvipdfmx]{graphicx}
\usepackage{braket}
\usepackage{amssymb, amsmath}
\usepackage[samesize]{cancel}
\usepackage{bm}
\usepackage{url}
\usepackage{jcappub}

\newcommand{\be}{\begin{eqnarray}}
\newcommand{\ee}{\end{eqnarray}}

\newcommand{\beq}{\begin{equation}}
\newcommand{\eeq}{\end{equation}}
\newcommand{\beqa}{\begin{eqnarray}}
\newcommand{\eeqa}{\end{eqnarray}}

\newcommand{\lkk}{\left[}
\newcommand{\rkk}{\right]}

\title{\boldmath Particle production induced by vacuum decay in real time dynamics}

\author[a,b]{Soichiro Hashiba,}
\author[a]{Yusuke Yamada,}
\author[a,b,c,d]{and Jun'ichi Yokoyama}

\affiliation[a]{Research Center for the Early Universe (RESCEU), Graduate School of Science,\\ The University of Tokyo, Hongo 7-3-1,
Bunkyo-ku, Tokyo 113-0033, Japan}
\affiliation[b]{Department of Physics, Graduate School of Science,
The University of Tokyo,\\ Hongo 7-3-1
Bunkyo-ku, Tokyo 113-0033, Japan}

\affiliation[c]{Kavli Institute for the Physics and Mathematics of the Universe (Kavli
 IPMU), UTIAS, WPI, The University of Tokyo, Kashiwa, Chiba, 277-8568, Japan}

\affiliation[d]{Trans-scale Quantum Science Institute, 
The University of Tokyo,\\ Hongo 7-3-1,
Bunkyo-ku, Tokyo 113-0033, Japan}

\emailAdd{sou16.hashiba@resceu.s.u-tokyo.ac.jp}
\emailAdd{yamada@resceu.s.u-tokyo.ac.jp}
\emailAdd{yokoyama@resceu.s.u-tokyo.ac.jp}

\subheader{{\rm RESCEU-11/20}}

\abstract{We discuss particle production associated with vacuum decay, which changes the mass of a scalar field coupled to a background field which induces the decay. By utilizing the Stokes phenomenon, we can optimally track the time-evolution of mode function and hence calculate particle production properly. In particular, we use real time formalisms for vacuum decay in Minkowski and de Sitter spacetime together with the Stokes phenomenon method. For each case, we consider the flyover vacuum decay model and stochastic inflation, respectively. Within the real time formalism, the particle production can be viewed as that caused by nontrivial external fields. This gives us a novel perspective of the real time formalism of vacuum decay.}
\makeatletter
\gdef\@fpheader{}
\makeatother

\begin{document}
\maketitle
\flushbottom
\section{Introduction}
Quantum fields coupled to a time-dependent background appear in various contexts, such as cosmology or more general curved spacetime. 
On such backgrounds, the definition of the vacuum state is not unique, and  change of vacuum states results in production of corresponding particles. Examples of such particle production are the Schwinger effect by the electromagnetic field~\cite{Heisenberg:1935qt,Schwinger:1951nm}, the Hawking radiation in the black hole spacetime~\cite{Hawking:1974rv} and the gravitational particle creation by change of the expansion law of the universe~\cite{Parker:1969au,Zeldovich:1971mw}. The efficiency of such particle production strongly depends on how abruptly the background changes~\cite{Ford:1986sy}. In the case of the gravitational particle creation, the transition time scale of the background metric determines 
typical energy scale of the produced particle~\cite{Hashiba:2018iff,Hashiba:2018tbu}. This implies that the a sudden transition of background makes the effect of particle production efficient. 

The vacuum decay such as a first-order phase transition is an example of such an abrupt transition as various parameters  change discontinuously. If  particle production associated with quantum tunneling is efficient, the tunneling dynamics could be affected by the back reaction of particle production. One of the most important possibilities is the issue of the Higgs field instability~\cite{EliasMiro:2011aa}. Since the Higgs field couples to almost all particles in the Standard Model, its tunneling might yield a considerable amount of the Standard Model particles. If this is the case, it would be necessary to reconsider the stability as well as the dynamics of the vacuum bubble after its nucleation.

Production of particles that are coupled to a tunneling scalar field has been studied in~\cite{Rubakov:1984pa,Yamamoto:1994te} using a conventional instanton method~\cite{Coleman:1977py,Callan:1977pt}. Despite our naive expectation of a sudden transition, it has been claimed that particle production is not so efficient: for a momentum $k$-mode, the number density of produced particles $n_k$ is exponentially suppressed for modes with energy $\omega_k\gg \Delta \tau^{-1}$ as $n_k \sim e^{-4\omega_k \Delta \tau}$. Here $\Delta \tau$ is an ``{\it imaginary} transition time scale", that is, the Euclidean time scale for the tunneling scalar moving from the false vacuum to the true one. One can find a similarity between this suppression factor and the one for the gravitational particle production case~\cite{Hashiba:2018iff,Hashiba:2018tbu}, in which $n_k$ is exponentially suppressed by $e^{-4\omega_k \Delta t}$, where $\Delta t$ is a real transition time scale. However, the meaning of the imaginary transition time scale is not clear from the {\it real time} perspective.\footnote{In~\cite{Andreassen:2016cff,Andreassen:2016cvx}, the vacuum transition rate is given without the notion of instanton and they give the meaning of tunneling rate within a real time formalism.}

In this paper, in order to understand how the particle production associated with the vacuum decay from real time viewpoint, we analyze the vacuum decay with its real time formulations. There are several description of the vacuum decay alternative to the standard Euclidean methods: In~\cite{Hertzberg:2019wgx}, the Schwinger-Keldysh formalism with Wigner function method is used to describe the quantum tunneling in quantum field theory.\footnote{In~\cite{Calzetta:2001gv,Calzetta:2001pp}, using the similar description, the effect of short wavelength modes on the long wavelength modes is discussed, which is called ``activation" and gives rise to enhancement of the vacuum decay rate. In this sense, we have to be aware that what the real time formalism actually describes might be this ``activation" rather than the very quantum tunneling in usual context.} In~\cite{Yasue:1978zz,PhysRevLett.40.1473,Braden:2018tky,Blanco-Pillado:2019xny,Huang:2020bzb}, a stochastic description of the quantum tunneling is discussed.\footnote{We should notice that there are considerable differences among the methods in~\cite{Yasue:1978zz,PhysRevLett.40.1473,Braden:2018tky,Blanco-Pillado:2019xny,Huang:2020bzb}. In~\cite{Yasue:1978zz,PhysRevLett.40.1473}, they take a stochastic quantum noise into account throughout the tunneling process with the Madelung fluid description~\cite{Nelson:1966sp,Guerra:1973ck}, whilst a stochastic quantum noise is used only as an initial kick and a succeeding dynamics is described classically without any stochastic component in~\cite{Braden:2018tky,Blanco-Pillado:2019xny,Huang:2020bzb}.} In the de Sitter spacetime, the stochastic inflation~\cite{Starobinsky:1986fx,Starobinsky:1994bd} can give the time dependent probability distribution of a scalar field value, which can actually give the tunneling rate corresponding to that found in the standard instanton methods. Such ``real time" formulations of the quantum tunneling are suitable for our purpose. However, we should emphasize that the initial state for the tunneling field seems different from that in the Euclidean description of the quantum tunneling. In particular, the ``flyover" vacuum decay which we will use in the flat spacetime case requires certain initial distribution for momentum of the tunneling field.  Nevertheless, the real time formalism can describe ``vacuum decay" and give the decay rate similar to that in instanton methods. Therefore, we simply call such vacuum decay as quantum tunneling in this work.

On top of such real time formulations, we will use the Stokes phenomenon method~\cite{Dingle:1973,Barry:1989zz,Dumlu:2010ua,Dumlu:2011rr}, which enables us to pick up a non-perturbative particle production, and also discuss how we can interpret the particle production caused by the vacuum decay in real time formulation.

This paper is organized as follows. In Sec.~\ref{sec:Stokes}, we give a brief review of the relation between particle production and the Stokes phenomenon. Such a viewpoint enables us to find the optimal evaluation of particle production in non-trivial backgrounds. In Sec.~\ref{sec:flat}, the production of a scalar particle coupling with a transiting scalar field in a flat spacetime background is investigated. First, we briefly review the real time formalism of the quantum tunneling and then evaluate a produced particle number density within the real time formalism. In Sec.~\ref{sec:dS}, we consider the particle production in the de Sitter spacetime background. After revisiting the particle production in the de Sitter spacetime, we extend our analyses to the case of a scalar field that couples to a transiting scalar field. In Sec.~\ref{sec:coda}, we summarize our results and discuss the remaining issues.

We use the natural units $c = \hbar = M_{\rm pl}$
throughout the paper.

\section{Particle production as Stokes phenomenon} \label{sec:Stokes}
In this section, we briefly review how particle production is interpreted in terms of the Stokes phenomenon~\cite{Dingle:1973,Barry:1989zz,Dumlu:2010ua,Dumlu:2011rr}. Particle production caused by time-dependent background can be understood from the behavior of mode functions. For particles in nontrivial time-dependent background, WKB type (adiabatic) mode functions are useful in defining the vacuum state. Such adiabatic solutions show sudden change of their behavior at a certain point, which is the so-called Stokes phenomenon. The physical meaning of this sudden change is nothing but the production of particles.

Let us consider a scalar field $\chi$ with a time-dependent mass $M_\chi^2(t)$ in the Minkowski background $ds^2 = -dt^2+d{\mathbf x}^2$. $\chi$ can be expanded as
\begin{equation}
    \chi(t, {\bf x})=\int \frac{d^3k}{(2\pi)^{3/2}}\left(\hat{a}_{\bf k}v_k(t)e^{{\rm i}{\bf k}\cdot{\bf x}}+
    \hat{a}_{\bf k}^{\dagger}\bar{v}_k(t)e^{-{\rm i}{\bf k}\cdot{\bf x}}\right),\label{ad}
\end{equation}
where we have introduced annihilation and creation operators $\hat{a}_{\bf k}$ and $\hat{a}_{\bf k}^{\dagger}$, respectively. The mode equation of $\chi$ is given by
\begin{equation}
    \ddot{v}_k+\omega_k^2v_k=0, \label{eomchi}
\end{equation}
where a dot denotes a time derivative and
\begin{equation}
    \omega_k^2 \equiv k^2 + M_\chi^2(t)
\end{equation}
is the effective frequency squared. 
In order to define a vacuum state, we take the WKB-type adiabatic solution for (\ref{eomchi}),
\begin{equation}
    v_k = \frac{A_k}{\sqrt{2W_k(t)}}e^{-{\rm i}\int^t dt'W_k(t')}+\frac{B_k}{\sqrt{2W_k(t)}}e^{{\rm i}\int^t dt'W_k(t')}, \label{WKBsoln}
\end{equation}
where $W_k$ is recursively determined as
\begin{align}
    W_k^{(0)} &= \omega_k, \\
    \left(W_k^{(n+1)}\right)^2 &= \omega_k^2 - \frac12 \left[ \frac{\ddot{W}_k^{(n)}}{W_k^{(n)}} - \frac32 \left( \frac{\dot{W}_k^{(n)}}{W_k^{(n)}} \right)^2 \right],
\end{align}
where $ W_k^{(n)}$ denotes the quantity of $n$-th adiabatic order~\cite{Dumlu:2011rr,Dabrowski:2014ica}.
We should stress that this $W_k^{(n)}$ is a divergent series. Therefore, we must truncate the expansion at the optimal order. The asymptotic behavior of such a WKB-type solution can be well described by a formula derived by Berry~\cite{Barry:1989zz}: The approximated adiabatic solution is multivalued function having the cut near the so-called turning points $t_c$ satisfying $\omega_k(t_c)=0$, and such turning point can be on a complex $t$-plane and usually associated with a complex conjugate point due to Schwarz's reflection principle. Hereafter, we denote the turning point closest to the real time axis in the upper half plane as $t_c$. The behavior of the solution significantly changes beyond the Stokes lines connecting a pair of turning points. More specifically, the values of the $\alpha_k$ and $\beta_k$ suddenly change --- in other words, particles are produced --- around the Stokes line. The Stokes phenomenon determines which order $n$ is optimal to be truncated. After the truncation, (\ref{WKBsoln}) becomes
\begin{equation}
    v_k = \frac{\alpha_k(t)}{\sqrt{2\omega_k(t)}}e^{-{\rm i}\int_{t_c}^t dt'\omega_k(t')}+\frac{\beta_k(t)}{\sqrt{2\omega_k(t)}}e^{{\rm i}\int_{t_c}^t dt'\omega_k(t')}. \label{univ}
\end{equation}
The functions $\alpha_k(t)$ and $\beta_k(t)$ is generalization of the Bogoliubov coefficients and $\beta_k(t)$ is responsible for particle production, which is given by~\cite{Barry:1989zz}
\begin{equation}
    \beta_k(t)\sim-{\rm i}e^{\frac12 F_k(t_c^*)}S_k(t), \label{beta}
\end{equation}
where $F_k(t)$, called a singulant, is given by
\begin{equation}
F_k(t) = 2{\rm i}\int_{t_c}^t dt' \omega_k(t'). \label{singulant}
\end{equation}
Here we take the integration contour along the Stokes line connecting $t_c$ and $t_c^*$ until the contour crosses the real axes at $t=t_s$, and from $t_s$ the contour is along the real axes. $S_k(t)$, called the Stokes multiplier, is given by
\begin{equation}
S_k(t)=\frac12\left[1+{\rm Erf}\left(-\frac{{\rm Im}F_k(t)}{\sqrt{2|{\rm Re}F_k(t)|}}\right)\right],
\end{equation}
and ${\rm Erf}(x)$ denotes the error function.
The Stokes line is the trajectory where the imaginary parts of the two exponents in (\ref{univ}) coincide: $\mathrm{Im}\:F_k(t) = 0$. In terms of particle production, the singulant represents the amplitude of particle production, whereas the Stokes multiplier represents the time dependence of the particle number. We refer to appendix A of~\cite{Li:2019ves} for a review of the derivation of this formula. In case there are $N$ pairs of turning points, a resultant particle number $n_k$ is given by summing up the contributions from each Stokes lines with relative phases as~\cite{Dumlu:2010ua}
\begin{equation}
  n_k=  |\beta_k(\infty)|^2 \sim \left| \sum_{n=0}^{N-1} \exp\left( 2{\rm i}\int_{t_{s,0}}^{t_{s,n}}\omega_k dt \right) e^{\frac12 F_{k,n}(t_{c,n}^*)} \right|^2, \label{Stokesbeta}
\end{equation}
where $t_{c,n}$, $t_{s,n}$ and $F_{k,n}$ denote the $n$-th turning point, the $n$-th intersection point between the $n$-th Stokes line and the real time axis, and the singulant with respect to $t_{c,n}$, respectively. The approximated formula \eqref{univ} gives the optimal approximation for the WKB type solutions, and therefore for the particle production rate. We will use this description of particle production in the following discussions.

\section{Flat spacetime case} \label{sec:flat}
In this section, we consider the production of a scalar field $\chi$ coupled to another scalar field $\phi$ that induces a quantum tunneling from a false vacuum to the true one in  flat spacetime background.

\begin{figure}[htbp]
	\centering
	\includegraphics[width=.64\textwidth]{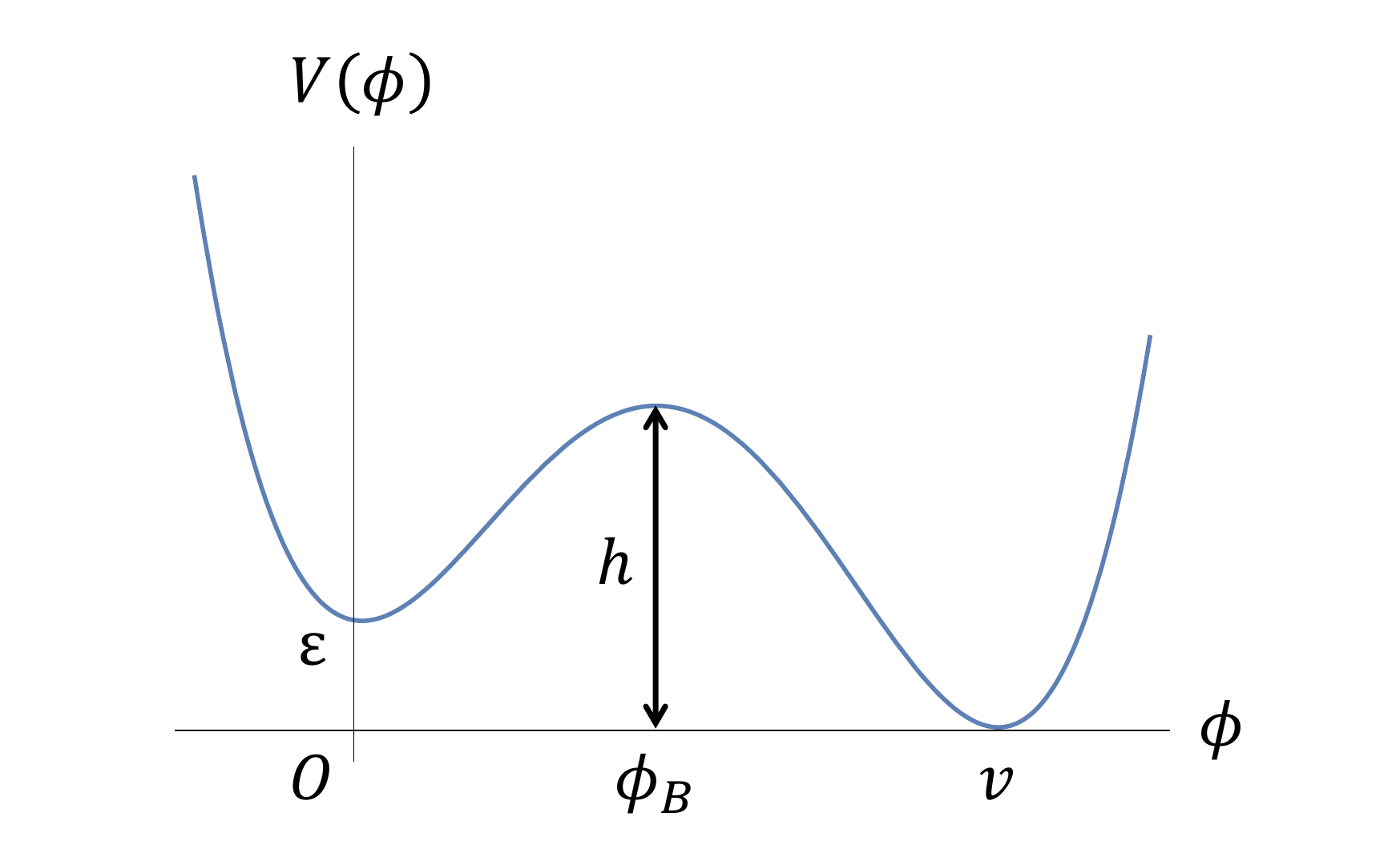}
	\caption{\label{fig:potential} A potential shape of a field $\phi$. Initially, $\phi$ is located at the false vacuum $\phi = 0$ and then it tunnels into the true vacuum $\phi = v$.}
\end{figure}

In the following discussion, we will discuss particle production associated with the flyover vacuum decay~\cite{Blanco-Pillado:2019xny}.
Let us consider the following system of a real scalar field $\phi$
\begin{equation}
    \mathcal{L} = -\frac12 \partial^\mu\phi \partial_\mu\phi - V(\phi).
\end{equation}
We assume the scalar potential $V(\phi)$ which has false and true vacua as shown in Fig.~\ref{fig:potential}.
Initially, $\phi$ sits at the false vacuum at $\phi=0$ and we consider the decay of the false vacuum to the true one at $\phi=v$. In the standard description of the vacuum decay, people use the Euclidean method~\cite{Coleman:1977py,Callan:1977pt}. Recently, a different description of the vacuum decay was proposed in~\cite{Blanco-Pillado:2019xny}, which is called the flyover vacuum decay. In this formalism, the tunneling field $\phi$ has an initial velocity fluctuation with its distribution given by
\begin{equation}
P(\dot{\phi}_0)={\cal N}\exp \left(-\frac{\dot{\phi}_0^2}{16p^2}\right).\label{inidis}
\end{equation}
where $\dot{\phi}_0$ is the initial velocity and ${\cal N}$ is a normalization factor. Here, we will briefly review the flyover model and how this Gaussian width $p$ is determined. First, we consider a velocity field operator smeared over a spherical region with radius $l$,
\begin{equation}
  \dot{\hat{\phi}}_l(t)=(2\pi l^2)^{-3/2}\int d^3{\bf x'}\dot{\hat{\phi}}({\bf x},t)\exp\left(-\frac{|{\bf x}-{\bf x}'|^2}{2l^2}\right).
\end{equation} 
 We assume the initial profile of the velocity field to take the spherically symmetric Gaussian form $\dot{\phi}({\bf x},t_0)=\dot{\phi}_0e^{-\frac{r^2}{2l^2}}$, as a boundary condition corresponding to the initial state of the velocity at $t=t_0$. Here $\dot{\phi}_0$ represents the initial velocity value of which distribution will be given by a Gaussian form as shown below.\footnote{We should think of the initial boundary condition imposed here are for  some finite region larger than the characteristic scale $l$. In the region we are considering,  $\phi$ is assumed to take a value around the false vacuum $\phi(t_0,{\bf x})= 0$ on average. The boundary condition for $\phi$ and $\dot\phi$ is different from that adopted in Euclidean method which does not specify the initial condition of $\dot{\phi}$.} Then the initial value of the smeared velocity is related to $\dot\phi_0$ as $\dot{\phi}_l(t_0)=2^{-3/2}\dot{\phi}_0$. Assuming that we may treat the scalar $\phi$ as a free field around the false vacuum, the distribution function of the smeared velocity is given by a Gaussian form
\begin{equation}
    P(\dot{\phi}_l)=\mathcal{N}\exp\left(-\frac{\dot{\phi}_l^2}{2\langle\dot{\phi}_l^2\rangle}\right)=\mathcal{N}\exp\left(-\frac{\dot{\phi}_0^2}{16\langle\dot{\phi}_l^2\rangle}\right),
\end{equation}
where the expectation value $\displaystyle \langle\dot{\phi}_l^2\rangle\simeq\frac{m}{16\pi^{3/2}l^3}$ is evaluated with the standard Minkowski vacuum state (see Appendix A of \cite{Blanco-Pillado:2019xny}).
In this model, the bubble nucleation can be understood as that the scalar field $\phi$ localized at the false vacuum has some probability to acquire an initial velocity enough to flyover the potential barrier. The most probable case is that the smeared initial velocity field barely exceeds the threshold value. Besides that, the bubble needs to expand after nucleation, which requires the minimal length scale of fluctuation to be $l\sim \frac{2\sigma}{\epsilon}$ where $\sigma\sim \int_0^v \sqrt{V}d\phi$ is the tension of the bubble, and $\epsilon$ is the difference of vacuum energy density between false and true vacua, which is $\epsilon=V(0)$ in our case. We should emphasize that the smearing scale $l$ is chosen to discuss the probability of finding a bubble expanding after nucleation, and therefore, the resultant probability distribution is not an ad hoc choice. 

The authors of \cite{Blanco-Pillado:2019xny} numerically show that the tunneling region in this formalism behaves as the standard picture of the bubble expansion, which is realized in the standard bounce calculation. We should note that the resultant decay rate is qualitatively the same as the standard instanton calculation, but the exponent of the decay rate is different from it. Therefore, strictly speaking, we should distinguish this model from the standard vacuum decay calculated by instanton methods. Nevertheless, we may think of it as a real time realization of the vacuum decay/tunneling.

Let us rewrite the probability distribution of the initial velocity. The critical bubble satisfies
\begin{equation}
    4\pi r_b^2 \sigma = \frac43 \pi r_b^3 \varepsilon,
\end{equation}
where $r_b$ is the critical bubble radius. From this relation, we find
\begin{equation}
    P(\dot{\phi}_0)=\mathcal{N}\exp\left[-\frac{2V_B}{9m_F}\dot{\phi}_0^2\right],\label{W0}
\end{equation}
where $m_F=\sqrt{V''(0)}$ is the mass of $\phi$ at the false vacuum and $V_B$ denotes the volume of the critical bubble $V_B=\frac{4\pi}{3}r_b^3$. For later convenience, we note that $\sigma$ is approximated as $v\sqrt h$, so that  $r_b = 3v\sqrt h/\varepsilon$.  Here $h$ is the height of the potential barrier. We will use the initial velocity distribution in evaluating the produced particle number density.

Then, we introduce an additional real scalar field $\chi$, which is coupled to the tunneling field $\phi$ and  described by the following Lagrangian
\begin{align}
\mathcal{L} = - \frac12 \partial^\mu\chi \partial_\mu\chi - \frac12(M_0^2 +g\phi^2)\chi^2.
\end{align}
Again, the potential $V(\phi)$ is a slightly tilted double-well shown in Fig.~\ref{fig:potential}, and $\chi$ is a heavy scalar field that acquires an additional mass $g\phi^2$ through tunneling. 
In this case, the effective frequency of $\chi$ is given by
\begin{equation}
    \omega_k^2 = k^2 + M_0^2 + g\phi^2 \label{flatomega}
\end{equation}
We assume that $g$ is sufficiently small and the tunneling dynamics is not affected by the coupling between $\phi$ and $\chi$. The coupling to the tunneling field with a particular initial velocity causes the time varying mass for the coupled scalar $\chi$. Under the assumption that $\chi$ does not affect the tunneling field dynamics, we are able to consider the tunneling field to be a classical background. Given an initial velocity $\dot{\phi}_0$ and an initial field value $\phi\sim 0$, we can estimate the time variation of $\omega_k^2$ in \eqref{flatomega} using the method reviewed above. Let us call the resultant number density of $\chi$ as $n_k(\dot{\phi}_0)$. Here we have explicitly shown the dependence on the initial velocity $\dot{\phi}_0$. Since the initial velocity obeys the Gaussian distribution~\eqref{inidis} and we are interested in the particle production when the bubble nucleation takes place, the particle number is given by the expectation value
\begin{equation}
\langle n_k\rangle=\int_{\dot{\phi}_{\rm th}}^\infty d\dot{\phi}_0 n_k(\dot{\phi}_0) P(\dot{\phi}_0),\label{averagenumber}
\end{equation}
where $\dot{\phi}_{\rm th}$ denotes the minimal velocity for bubble nucleation to take place. 

Now we move onto the evaluation of produced particle number density. As we have discussed in Sec.~\ref{sec:Stokes}, we can evaluate the $\chi$-production by considering the behaviour of the mode function of $\chi$ with the background external field $\phi$. Given that $\phi$ acquires an initial velocity $\dot{\phi}_0$ at $t = t_0$ inside a spherical region $S_B$ with the radius $r_b$, $\phi$ is approximately homogeneous inside $S_B$ and therefore the gradient energy of $\phi$ is negligible compared with the kinetic energy. Since it is known that the particle production efficiently occurs at the point where non-adiabaticity parameter $\dot{\omega}_k/\omega_k^2$ takes the local maximum (see e.g.~\cite{Kofman:1997yn}), we may assume that the $\chi$-production occurs almost only when $\dot\phi$ is large, namely around each vacuum. (We will verify this statement by analyzing the Stokes lines in Appendix~\ref{app}.) We expand $V(\phi)$ around each local minimum as
\begin{equation}
    V(\phi) = \begin{cases}
        \varepsilon + \frac{1}{2} m_F^2 \phi^2 + \mathcal{O}(\phi^3) & (\phi \approx 0) \\
        \frac{1}{2} m_T^2 (\phi - v)^2 + \mathcal{O}((\phi - v)^3) & (\phi \approx v)
    \end{cases},
\end{equation}
where $m_T^2 = V''(v)$, and the motion of $\phi$ inside $S_B$ is approximated, respectively, as
\begin{equation}
    \phi(t) \approx \begin{cases}
        \frac{\dot{\phi}_0}{m_F} \sin\left[ m_F (t - t_0) \right] & (\phi \approx 0) \\
        \frac{\sqrt{\dot{\phi}_0^2 + 2\varepsilon}}{m_T} \sin\left[ m_T (t - t_2) \right]+v & (\phi \approx v)
    \end{cases}. \label{phiapprox}
\end{equation}
Here we assume that the particle production of $\chi$ is so inefficient that the energy of $\phi$ is almost conserved, which is actually the case as we will see. 

Let us find the turning points and the accompanying Stokes line for the region near each vacuum. Around the false vacuum $\phi \approx 0$, the turning point $t_{c,0}$ is obtained by substituting the approximated expression \eqref{phiapprox} into the equation $\omega_k = 0$ as
\begin{equation}
    t_{c,0} = t_0 + {\rm i}m_F^{-1} \sinh^{-1}x,
\end{equation}
where $x = \sqrt{\frac{m_F^2}{g \dot{\phi}_0^2} (k^2 + M_0^2)}$ and $\sinh^{-1}x = \ln\left( x + \sqrt{1 + x^2} \right)$. The singulant along the Stokes line connecting the pair of these turning points is given by
\begin{align}
    F_{k,0}(t_{c,0}^*) &\approx 2{\rm i}\int_{t_{c,0}}^{t_{c,0}^*} \sqrt{k^2 + M_0^2 + \frac{g\dot\phi_0^2}{m_F^2}\sin^2[m_F (t-t_0)]} dt \nonumber \\
    &= 4\int_0^{\sinh^{-1}x} \sqrt{\frac{g\dot\phi_0^2}{m_F^2}(x^2 - \sinh^2 \xi)} m_F^{-1} d\xi \qquad ({\rm i}\xi = m_F(t-t_0)) \nonumber \\
    &= 4\sqrt\frac{g\dot\phi_0^2}{m_F^4} (-{\rm i}x){\rm E}\bigg({\rm i}\sinh^{-1}x \bigg| -\frac{1}{x^2} \bigg), \label{Fk0raw}
\end{align}
where ${\rm E}(\varphi|k^2)$ is the incomplete elliptic integral of the second kind in trigonometric form. Although it is difficult to estimate the value of this function in general, we may use the approximated form for $x \gg 1$ since we are considering a heavy original mass $M_0$ and a small coupling $g$. For $x \gg 1$, we may expand the integral in $1/x$ and find the leading order of \eqref{Fk0raw} to be
\begin{equation}
    F_{k,0}(t_{c,0}^*) \simeq 4\sqrt\frac{g\dot\phi_0^2}{m_F^4} x [\ln(4x) - 1]. \label{Fk0}
\end{equation}
Next, let us evaluate the particle production around the true vacuum. Near the true vacuum $\phi \approx v$, the turning point is located at
\begin{align}
    t_{c,2} &= t_2 + m_T^{-1} \sin^{-1}\left[ \sqrt{\frac{m_T^2}{g(\dot{\phi}_0^2 + 2\varepsilon)}} \left(-\sqrt{gv^2} + {\rm i} \sqrt{k^2 + M_0^2}\right) \right] \\
    &= t_2 + {\rm i}m_T^{-1} \ln\left( z + \sqrt{1 + z^2} \right) - \frac{m_T^{-1}}{\sqrt{1 + z^2}} \sqrt\frac{g v^2}{k^2 + M_0^2} + \mathcal{O}\left( z^{-2}\frac{g v^2}{k^2 + M_0^2} \right),
\end{align}
where $z = \sqrt{\frac{m_T^2}{g(\dot{\phi}_0^2 + 2\varepsilon)}(k^2 + M_0^2)}$. Here we have shown the leading order terms in $\delta \equiv \sqrt\frac{g v^2}{k^2 + M_0^2} \ll 1$. This approximation is valid as long as the generated mass $gv^2$ is sufficiently smaller than the original mass $M_0^2$. If this is not the case, $\chi$ particles are hardly produced around the true vacuum because the required energy for the $\chi$-production is too large. (See also Appendix~\ref{app}.) The singulant along the Stokes line connecting the pair of these turning points is given by
\begin{align}
    F_{k,2}(t_{c,2}^*) &= 2{\rm i}\int_{t_{c,2}}^{t_{c,2}^*} \sqrt{k^2 + M_0^2 + g\Bigg( \frac{\sqrt{\dot\phi_0^2 + 2\varepsilon}}{m_T}\sin[m_T (t-t_2)] + v \Bigg)^2} dt. \label{Fk2before}
\end{align}
In general, analytic calculation of this integral is quite difficult, but in a particular case, we can evaluate it semi-analytically. Here, we consider the double-well potential $V(\phi) \approx \frac{\lambda}{2} \phi^2 (\phi - v)^2$, and look for the expression in the case of the minimal velocity for bubble nucleation $\dot\phi_{\rm th}^2 = \lambda v^4/32$, which is only relevant for later discussion (see the discussion below). Then, \eqref{Fk2before} is simplified as
\begin{align}
    F_{k,2}(t_{c,2}^*) &= 2{\rm i}\int_{t_{c,2}}^{t_{c,2}^*} \sqrt{k^2 + M_0^2 + gv^2 \left( \frac{1}{\sqrt{32}}\sin\left[\sqrt{\lambda}v (t-t_2)\right] + 1 \right)^2} dt \nonumber \\
    &= 2\delta^{-1} \int_{-\sqrt{32}/\delta}^{\sqrt{32}/\delta} \sqrt\frac{1 - \delta^2 X^2/32}{1 + (X - \sqrt{32}{\rm i})^2} dX \sqrt\frac{g}{\lambda}, \label{Fk2b2}
\end{align}
where $X = {\rm i}\left( \sin[m_T (t-t_2)] + \sqrt{32} \right)$. We numerically find a fitting function of \eqref{Fk2b2} to be
\begin{align}
    F_{k,2}(t_{c,2}^*) \simeq \mathcal{D} \delta^{-(1+\mathcal{E})} \sqrt\frac{g}{\lambda}, \label{Fk2}
\end{align}
where $\mathcal{D} \approx 10.5$ and $\mathcal{E} \approx 0.13$ are numerical coefficients.

Combining the particle production around each vacuum, we obtain the resultant number density of produced $\chi$ from \eqref{Stokesbeta} as
\begin{align}
    \langle n_k(\dot\phi_0)\rangle = \left| e^{-\frac12 F_{k,0}(t_{c,0}^*)} + \exp\left( 2{\rm i}\int_{t_{s,0}}^{t_{s,2}}\omega_k(t)dt \right) e^{-\frac12 F_{k,2}(t_{c,2}^*)} \right|^2.
\end{align}
Since $t_{s,2} - t_{s,0}$ is much larger than $\omega_k^{-1}$ in the situation where $\phi$ barely passes over the potential barrier, we can take average of the relative phase and obtain
\begin{align}
    \langle n_k(\dot\phi_0)\rangle &\approx e^{-F_{k,0}(t_{c,0}^*)} + e^{-F_{k,2}(t_{c,2}^*)}. \label{nkp0}
\end{align}
Finally, we calculate the expectation value of produced number density of $\chi$ by integration weighted by the distribution function. Substituting \eqref{W0} into \eqref{averagenumber}, we obtain
\begin{align}
    \langle n_k \rangle &= \int_{\dot\phi_{\rm th}}^\infty d\dot\phi_0 \sqrt{\frac{2V_B}{9\pi^{1/2}m_F}} \exp\left[ -\frac{2\pi^{1/2}V_B}{9m_F}\dot{\phi}_0^2 \right]\langle n_k(\dot\phi_0)\rangle, \label{praiseYY}
\end{align}
where $\dot\phi_{\rm th}^2 \equiv 2(h-\varepsilon)$ is the initial energy threshold requisite for flying over the barrier. Here, the Gaussian factor of the distribution function rapidly decreases for larger $\dot\phi_0$, and thus most of contribution of this integration comes from around $\dot\phi_0 \approx \dot\phi_{\rm th}$. Therefore, we can approximate \eqref{praiseYY} as
\begin{align}
    \langle n_k \rangle &\approx \int_{\dot\phi_{\rm th}}^\infty d\dot\phi_0 \sqrt{\frac{2V_B}{9\pi^{1/2}m_F}} \exp\left[ -\frac{2\pi^{1/2}V_B}{9m_F}\dot{\phi}_0^2 \right] \langle n_k(\dot\phi_{\rm th})\rangle \nonumber \\
    &= \frac12 {\rm Erfc}\left(\sqrt{\frac{2\pi^{1/2}V_B}{9m_F}}\dot{\phi}_{\rm th}\right) \langle n_k(\dot\phi_{\rm th})\rangle,
\end{align}
where ${\rm Erfc}(x)=\frac{2}{\sqrt{\pi}} \int^\infty_xe^{-t^2}dt$ is the complementary error function. Using the asymptotic form of ${\rm Erfc}(x)$ given by
\begin{equation}
    {\rm Erfc}(x)\sim \frac{e^{-x^2}}{x\sqrt\pi} \label{asympErfc}
\end{equation}
for $x \gg 1$ together with \eqref{Fk0}, \eqref{Fk2} and \eqref{nkp0}, we can further proceed analytic calculation as
\begin{align}
    \langle n_k \rangle &\approx \sqrt{\frac{9m_F}{2\pi^{3/2}V_B\dot{\phi}_{\rm th}^2}}\exp\left[ -\frac{2\pi^{1/2}V_B}{9m_F}\dot{\phi}_{\rm th}^2 \right] \nonumber \\
    &\qquad \times \left( \exp \left[ -2\sqrt\frac{k^2 + M_0^2}{\lambda v^2} \ln\left( \frac{512}{{\rm e}^2} \frac{k^2 + M_0^2}{gv^2} \right) \right] + \exp \left[ -\mathcal{D} \delta^{-(1+\mathcal{E})} \sqrt\frac{g}{\lambda} \right] \right). \label{rb_nchi}
\end{align}
Here, we have assumed the slightly tilted double-well potential $V(\phi) \approx \frac{\lambda}{2} \phi^2 (\phi - v)^2$ and then $m_F = m_T = m_\phi$ for simplicity.\footnote{Although the first term in the second line of \eqref{rb_nchi}, which comes from \eqref{Fk0}, seems to depend on $v$, this is simply because we assume the double-well potential and then $m_F^2 = \lambda v^2$ and $\dot\phi_{\rm th}^2 = \lambda v^4/32$ are satisfied.} If one would like to consider more general potential, some of our intermediate results can be applicable; however it would require more involved calculations. Noticing that the overall factor in the first line is nothing but the tunneling rate $\Gamma$, we obtain the following simpler expression:
\begin{equation}
    \langle n_k \rangle \approx \Gamma \left( \exp \left[ -2\Delta \ln\left( \frac{512}{{\rm e}^2} \Delta^2 \right) \sqrt\frac{g}{\lambda} \right] + \exp \left[ -\mathcal{D} \Delta^{1+\mathcal{E}} \sqrt\frac{g}{\lambda} \right] \right), \label{result_nchi}
\end{equation}
where $\Delta^2 \equiv (k^2 + M_0^2)/gv^2$. Since the produced particle number density decays exponentially or even faster for high momentum modes, the total particle number density safely converges.

Let us discuss the physical meaning of the two contributions for the produced particle number given in \eqref{result_nchi}. The first contribution coming from the Stokes line crossing $t_0$ is not the production due to the change of mass but that caused by the initial velocity of $\phi$. Such a contribution exists no matter how small or large the mass difference is. The second contribution corresponds to the production by the transition of the mass of $\chi$ which would be more intuitive than the first one. 

We should stress that we have only discussed the particle production from ``one-way" process, $\phi=0$ to $\phi=v$. This is because we have approximated the dynamics of the tunneling scalar $\phi$ to be homogeneous. Although this approximation makes analyses of particle production simpler, we find that the oscillation of the tunneling scalar will never stop. However, the bubble of true vacuum would form once $\phi$ reaches the true vacuum and then its wall would extract the energy of $\phi$, then $\phi$ will never oscillate between true and false vacuum \cite{Blanco-Pillado:2019xny,Huang:2020bzb}, which justifies the one-way process we have considered. Since the dominant effect for particle production comes from the homogeneous part inside the nucleating bubble, our estimation would be not so affected even if we take spatial dependence of $\phi$ into account.

Finally, let us compare our result with the one derived by using instanton methods. In~\cite{Rubakov:1984pa,Yamamoto:1994te}, it was found that the resultant particle spectrum shows thermal-like spectra with temperature given by the inverse of the imaginary time interval for scalar field dynamics in Euclidean space. On the other hand, as seen from \eqref{result_nchi}, our result does not show such a behavior. There are various reasons for the difference between our result and that in \cite{Rubakov:1984pa,Yamamoto:1994te}: Obviously, we have used the real time formalism, and notion of the Euclidean time does not show up in our description. Actually, our description of the tunneling is associated with the oscillating scalar field in real time whereas both the tunneling scalar and spectator scalar $\chi$ experience the Euclidean time evolution in the case of \cite{Rubakov:1984pa,Yamamoto:1994te}. In particular, in \cite{Yamamoto:1994te}, the authors used Milne and Rindler coordinates to discuss the particle production. As is well known, the Minkowski vacuum can be seen as thermal states in Rindler spacetime, and we suspect that the notion of the thermality in the resultant particle spectrum is related to such a coordinate system. In our discussion, we have used flat Minkowski coordinate, and such difference of the coordinate system may also lead to the difference of the spectra. It would be worth studying the tunneling dynamics in real Rindler/Milne coordinate system to discuss how the particle production associated with real time tunneling can be seen. This is beyond the scope of our paper, and we will leave such a question. We should also note that we have not taken into account spatial dependence of the wall dynamics as mentioned above. We expect it is not a crucial reason of the different spectrum because \cite{Rubakov:1984pa} also considers the homogeneous tunneling approximation but obtains the thermal-like particle spectrum. The difference of the particle spectrum may be thought of as the fundamental difference of the Euclidean and real time formalism. In order to discuss which is the correct description, we need better understanding of the real time tunneling description itself.~\footnote{As a different formalism of a real time tunneling description, the stochastic approach was investigated in~\cite{Goncharov:1986ua,Linde:1991sk}. Such a formalism might also be useful to understand the particle production associated with tunneling dynamics.}

\section{de Sitter spacetime case} \label{sec:dS}
In this section, we consider  production of a massive scalar particle $\chi$ coupled to a tunneling scalar $\phi$ in the de Sitter spacetime. In the vacuum decay in the de Sitter spacetime, there is an expansion effect as well as the dynamics of $\phi$, we will see that the particle production is rather different from that in the flat spacetime case.

\subsection{Particle production without tunneling dynamics}
For comparison with the later discussion, we start with the discussion on the particle production in the de Sitter spacetime without tunneling dynamics. We should stress that particle production without tunneling in the de Sitter spacetime is not physically expected. If we start with Bunch-Davies vacuum state, there would be no particle production since it is de Sitter invariant. Nevertheless, we demonstrate the ``particle production" by introducing an adiabatic vacuum because such a discussion is useful to understand the case with tunneling dynamics. For simplicity, we will consider a conformally coupled massive scalar field $\chi$. Here we have introduced the conformal coupling to simplify the form of the effective frequency $\omega_k$, but it is not essentially important. The system is described by the following Lagrangian:
\begin{equation}
\sqrt{-g}\mathcal{L}=-\frac12\sqrt{-g}\left(\partial_\mu\chi\partial^\mu\chi+\frac16R\chi^2+M^2\chi^2\right).
\end{equation}
Assuming the de Sitter background metric $ds^2=\frac{1}{H^2\eta^2}(-d\eta^2+d{\bf x}^2)$, we mode-expand $\chi$ as 
\begin{equation}
\chi=\int \frac{d^3k}{(2\pi)^{3/2}a(\eta)}\left(\hat{a}_{\bf k}v_k(\eta)e^{{\rm i}{\bf k}\cdot{\bf x}}+\hat{a}_{\bf k}^{\dagger}\bar{v}_k(\eta)e^{-{\rm i}{\bf k}\cdot{\bf x}}\right),
\end{equation}
where $a(\eta)=-\frac{1}{H\eta}$ is a scale factor, and we have introduced a creation (annihilation) operator $\hat{a}_{\bf k}$ ($\hat{a}_{\bf k}^{\dagger}$).
The mode equation of the scalar field is given by
\begin{equation}
v_k''+\omega_k^2v_k=0,
\end{equation}
where a prime denotes a derivative with respect to the conformal time $\eta$ and
\begin{equation}
    \omega_k^2(\eta) = k^2+\frac{M^2}{(-H\eta)^2}.
\end{equation}
Following the discussion in Sec.~\ref{sec:Stokes}, let us discuss the particle production in our case. The turning points where $\omega_k(\eta_c)=0$ in the complex $\eta$-plane, are simply given by
\begin{equation}
\eta_c=0 + {\rm i}\frac{M}{Hk}.
\end{equation}
The value of the singulant along the line connecting these turning points is given by
\begin{equation}
    \frac12 F_k(\eta_c^*) = {\rm i}\int_{\eta_c}^{\eta_c^*} d\eta'\omega_k(\eta') = {\rm i}\int_{\eta_c}^{\eta_c^*}d\eta' \sqrt{k^2 +\frac{M^2}{H^2\eta'^2}}.
\end{equation}
We integrate this along the path avoiding the simple pole $\eta'=0$, and only the half-pole integration contributes because other parts cancel. Parametrizing $\eta=\epsilon e^{{\rm i}\theta}$ ($\theta:\frac\pi 2\to \frac{3\pi}{2}$), we find
\begin{equation}
    \frac12 F_k(\eta_c^*) = -\lim_{\epsilon\to0}\int^{\frac{3}{2}\pi}_{\frac\pi2} e^{{\rm i}\theta}d\theta\sqrt{\epsilon^2k^2+\frac{M^2e^{-2{\rm i}\theta}}{H^2}} = -\frac{\pi M}{H}.
\end{equation}
Subsituting this into \eqref{Stokesbeta}, we obtain the asymptotic produced particle number as~~\cite{Mottola:1984ar,Garriga:1994bm}
\begin{equation}
n_k=|\beta_k|^2=e^{-2\pi M/H}. \label{ps}
\end{equation}
There are two problems in this analysis: One is that, since Stokes line is near the end point $\eta=0$, the mode function beyond the Stokes line might not be available. Therefore, this Bogoliubov coefficient might not have a clear physical meaning. The other is that $\beta_k$ does not depend on $k$, which seems to cause the infinite number of particle production, and the corresponding state is not normalizable and cannot be related to the original vacuum by unitary transformations. 

The former issue might be circumvented by using the coordinate time $t$ instead of the conformal time $\eta$. Let us use the coordinate system $ds^2=-dt^2+e^{2Ht}d{\bf x}^2$, and the mode equation of the scalar is given by~\cite{Li:2019ves}
\begin{equation}
    \ddot f_k+\omega^2_kf_k = 0,
\end{equation}
where we have parametrized $\chi$ as
\begin{equation}
    \chi(t,{\bf x})=\int \frac{d^3k}{(2\pi)^3}a^{-3/2}\left(a_{\bf k}f_k(t)e^{{\rm i}{\bf k}\cdot {\bf x}}+a^\dagger_{{\bf k}}\bar{f}_k(t)e^{{-\rm i}{\bf k}\cdot {\bf x}}\right)
\end{equation}
and
\begin{equation}
    \omega^2_k(t) = k^2e^{-2Ht}+M^2.
\end{equation}
The WKB solution to the mode equation takes the form
\begin{equation}
    f_k(t)= \frac{\alpha_k(t)}{\sqrt{\omega(t)}}e^{-{\rm i}\int^t dt' \omega_k(t')}+ \frac{\beta_k(t)}{\sqrt{\omega(t)}}e^{{\rm i}\int^t dt' \omega_k(t')}.
\end{equation}
Here we have chosen the WKB solution instead of the known exact solution that defines the Bunch-Davies vacuum, and our WKB solution corresponds to an adiabatic vacuum.\footnote{The choice of the vacuum in quantum field theory corresponds to the choice of the boundary conditions of mode functions. Since we solved the mode equation with WKB method, the boundary condition (= the choice of the vacuum) should be different from that of Bunch-Davies one.}
The turning point is given by
\begin{equation}
    t_c = -H^{-1}\left(\ln (M/k) + \frac{\pi}{2}{\rm i}\right).\label{tc}
\end{equation}
It is easy to derive the singulant along the Stokes line, and we find
\begin{equation}
    \frac12 F_k(t_c^*) = {\rm i}\int_{t_c}^{t_c^*} \omega_k(t')dt'=-\frac{M\pi}{H}
\end{equation}
Thus, the asymptotic Bogoliubov coefficient is 
\begin{equation}
\beta_k={\rm i}e^{-\frac{M\pi}{H}}\label{beta2}
\end{equation}
which is precisely the same as the one in \eqref{ps}. In this case, since the coordinate time $t$ varies from $t=-\infty$ to $t=\infty$, we expect that there exist the asymptotic mode function with the Bogoliubov coefficient~\eqref{beta2} for sufficiently large $t$. However, the Bogoliubov coefficients and therefore the particle number is independent of the momentum as is the case with conformal time, which causes the divergence. Therefore, the particle production caused by these turning point should not be realized physically; otherwise the late time vacuum cannot be related to the initial one via unitary transformation.

However, we should notice that we have so far discussed the behavior with comoving momenta. More careful treatment is necessary to discuss the particle production with physical momenta. The singulant at time $t$ is given by
\begin{equation}
F_k(t) = \frac{1}{2H} \left[ 2 {\rm i}e^{-H t} \sqrt{M^2 e^{2 H t}+k^2}-2 {\rm i} M \log \left(\frac{\sqrt{M^2 e^{2 H
   t}+k^2}+M e^{H t}}{k}\right)+\pi  M \right].
\end{equation}
Thus we find
\begin{align}
{\rm Re}\:F_k(t)=&\frac{\pi M}{2H},\\
{\rm Im}\:F_k(t)=&H^{-1}\left(\omega_k(t)-M\log\left(\frac{\omega_k(t)+M}{ke^{-Ht}}\right)\right),
\end{align}
with which the Stokes multiplier is given by
\begin{equation}
S_k(t)=\frac12\left(1+{\rm Erf}\left[-(\pi H M)^{-\frac12}\left(\omega_k(t)-M\log\left(\frac{\omega_k(t)+M}{ke^{-Ht}}\right)\right)\right]\right).
\end{equation} 
Therefore, the time-dependent Bogoliubov coefficient $\beta_k(t)$ is 
\begin{equation}
    \beta_k(t) = -\frac{{\rm i}e^{-\frac{M\pi}{H}}}{2} {\rm Erfc}[f(t)].\label{betak}
\end{equation}
where
\begin{equation}
    f(t) \equiv (\pi H M)^{-\frac12}\left(\omega_k(t)-M\log\left(\frac{\omega_k(t)+M}{ke^{-Ht}}\right)\right).
\end{equation}
and we have used the relation, ${\rm Erf}(-x)=-{\rm Erf}(x)$ and ${\rm Erf}(x)=1-{\rm Erfc}(x)$. In terms of the physical momentum $k_{\rm phys}=ke^{-Ht}$, the quantity $f$ is written as
\begin{equation}
    f(t) = (\pi H M)^{-\frac12}\left(\sqrt{k_{\rm phys}^2+M^2}-M\log\left(\frac{\sqrt{k_{\rm phys}^2+M^2}+M}{k_{\rm phys}}\right)\right),\label{fphys}
\end{equation}
which looks time independent. From this expression, we find (formally) time-independent particle spectrum produced in the de Sitter background. 
Since the asymptotic form of complementary error function is given by \eqref{asympErfc}  and $f\sim k_{\rm phys}/(\pi HM)^{\frac12}$, the particle number density for high physical momentum decays as 
\begin{equation}
n_{k_{\rm phys}}\sim \frac{HMe^{-\frac{2M\pi}{H}}}{4k_{\rm phys}^2}e^{-\frac{2k_{\rm phys}^2}{\pi H M}}.
\end{equation}
This would lead to a finite number of particle at any time $t$. The total number of the particle is
\begin{equation}
N_{\rm tot}=\int \frac{d^3k}{(2\pi)^3}n_k=e^{3Ht}\int\frac{d^3k_{\rm phys}}{(2\pi)^3}\frac{e^{-2\frac{M\pi}{H}}}{4}({\rm Erfc}(f))^2,
\end{equation}
where $f$ in terms of physical momentum is given in \eqref{fphys}. The integral would converge and give some finite value. For example, if we take $H=1,M=10$, numerical integration gives $N_{\rm tot}\sim 1.12\times 10^{-14} e^{3t}$. Therefore the total number of particle inside the comoving volume $N_{\rm tot}/a^{3}$ is finite.

We also note that the particle spectrum~\eqref{ps} is slightly different from the well-known results~\cite{Mottola:1984ar,Garriga:1994bm} given by
\begin{equation}
n_k^{dS}=\frac{1}{e^{2\pi M/H}-1}.
\end{equation}.This spectrum can be obtained by defining the (non-adiabatic) late time vacuum state and comparing it with the Bunch-Davies one.\footnote{Notion of particle production appears when we have two different vacuum states. In this case, the late time vacuum is chosen as the comparison state, which differs from the Bunch-Davies one. Since the late time mode function~\cite{Mottola:1984ar} has no momentum dependence and is not adiabatic. As a result, the spectrum cannot be convergent in integrating over momentum space as discussed below. As we mentioned earlier, if we only consider Bunch-Davies vacuum, there would be no notion of particle production because of the absence of the comparison state. Note that, however, the behavior of the mode function in early and late time has different asymptotic expansion, and one may regard the difference to be ``particle production". } This particle spectrum looks that of zero mode of a massive particle in thermal bath with $T=\frac{H}{2\pi}$, known as the Gibbons-Hawking temperature~\cite{Gibbons:1977mu}. Since we have discussed an adiabatic vacuum, the particle spectrum does not coincide with that of the Bunch-Davies vacuum. However, we find that the leading order is the same as \eqref{ps} in the super-massive limit $M/H\gg1$. In either way, this particle spectrum does not give the convergence of the momentum integral and in such a case, strictly speaking, the two vacuum states we are comparing cannot be transformed to each other via unitary transformation (see e.g. \cite{Parker:1969au}) and the notion of particle production does not really make sense there.

\subsection{Particle production with tunneling}
In the following, we discuss the particle production induced by the tunneling dynamics of the background scalar field. In this case, we would expect the shift of the turning point by the tunneling dynamics. Let us consider the following system:
\begin{equation}
\sqrt{-g}\mathcal{L}=-\frac12\sqrt{-g}\left(\partial_\mu\phi\partial^\mu\phi+V(\phi)+\partial_\mu\chi\partial^\mu\chi+\frac16R\chi^2+g\phi^2\chi^2\right),
\end{equation}
where $\phi$ denotes a real scalar field. $\phi$ is supposed to be the tunneling field, which initially sits at the false vacuum and eventually penetrates to the true vacuum. For our purpose, we apply the stochastic inflation formalism to the dynamics of $\phi$ at the zeroth order in $g$, which gives the coarse-grained dynamics of $\phi$. In the stochastic inflation formalism~\cite{Starobinsky:1986fx,Starobinsky:1994bd}, one integrates large momentum modes out, which yields stochastic noise for lower frequency modes being regarded as a classical field. Even though we average over super-horizon modes, namely average over different Hubble patches, the expectation value in a single patch would asymptote to the super-horizon average, as long as we are interested in sufficiently long time interval.

Particularly, we focus on the one-point probability distribution function $\rho[\phi(x)]$ obeying the following Fokker-Planck equation~\cite{Starobinsky:1994bd}
\begin{equation}
   \frac{ \partial}{\partial t}\rho[\phi(x)]=\frac{1}{3H}\frac{\partial}{\partial\phi}\{V'[\phi(x)]\rho[\phi(x)]\}+\frac{H^3}{8\pi^2}\frac{\partial^2}{\partial\phi^2}\rho[\phi(x)],
\end{equation}
where the prime denotes the functional derivative with respect to $\phi(x)$. The general solution of this Fokker-Planck equation is
\begin{equation}
\rho(\phi,t)=\exp\left(-\frac{4\pi^2V(\phi)}{3H^4}\right)\sum_{n=0}^\infty a_n\Phi_n(\phi)e^{-\Lambda_n(t-\tau)},\label{pdf}
\end{equation}
where $a_n$ is a constant and $\tau$ is the initial time, which we will take to be $\tau=-\infty$. Here, $\Phi_n(\phi)$ is the eigenfunction satisfying the following equation,
\begin{equation}
   \left[ -\frac{1}{2}\frac{\partial^2}{\partial\phi^2}+W(\phi)\right]\Phi_n(\phi)=\frac{4\pi^2\Lambda_n}{H^3}\Phi_n(\phi)
\end{equation}
where $\Lambda_n$ is a non-negative eigenvalue and
\begin{align}
    W(\phi)\equiv& \frac12 [v'(\phi)^2-v''(\phi)],\\
    v(\phi)\equiv &\frac{4\pi^2}{3H^4}V(\phi).
\end{align}
The lowest eigenvalue $\Lambda_0=0$ corresponds to the equilibrium mode, and we will take the mode up to the second lowest mode $n=1$. For the case with double well potential $V=\frac{\lambda}{4}(\phi^2-m^2/\lambda)^2$, we find 
\begin{equation}
    \Lambda_1=\frac{\sqrt{2}m^2}{3\pi H}\exp\left(-\frac{2\pi^2m^4}{3\lambda H^4}\right).
\end{equation}
Notice that the eigenvalue $\Lambda_1$ is exponentially suppressed, and $\Lambda_1\ll H$. Although the potential with false and true vacuum should not be exactly the same as the double well potential, the difference of the eigenvalues would not be so large. The time dependent expectation value of $\langle\phi^2(t)\rangle$ can be evaluated by the one-point probability distribution function. We discuss it in Appendix B, and here we simply give the result
\begin{equation}
\langle\phi^2 (t)\rangle=v^2(1-e^{-\Lambda (t-\tau)}),\label{vev}
\end{equation}
where $v^2$ denotes the vacuum expectation value of $\phi^2$ at the true vacuum, and we have assumed $\langle\phi\rangle=0$ at the false vacuum. If the explicit form is assumed, one is able to find the eigenvalues $\Lambda_n$ explicitly e.g. by perturbative methods.

The time dependent expectation value yields the time dependent mass term for the scalar $\chi$, from which the particle production takes place besides that caused by de Sitter background.
This expectation value becomes (infinitely) negative for $t\to-\infty$, which is not physically acceptable. Therefore, we may avoid such an issue by taking $\tau\to-\infty$. Instead of such a prescription, we will consider the following phenomenological modeling
\begin{equation}
    \langle\phi^2(t)\rangle_{\rm reg}=\frac12v^2\left(1+\tanh\left(\frac12 \Lambda t\right)\right),
\end{equation}
which asymptotically reproduce the original expression for $t>0$ while avoiding negative value of $\langle\phi^2\rangle$ for $t<0$. 

In this case, the frequency for a comoving momentum $k$ mode is given by
\begin{equation}
    \omega_k(t) = \sqrt{k^2e^{-2Ht}+\frac12 M^2\left(1+\tanh\left(\frac12\Lambda t\right)\right)}
\end{equation}
where $M^2\equiv gv^2$. Because of the complication of the frequency, it is impossible to find the analytic expression for the singulant $F_{k}(t)=2{\rm i}\int_{t_c}^t\omega_k(t')dt'$, although it would be possible to calculate it numerically. Therefore, we discuss the high and the low momentum modes separately with approximation. For the former case, according to \eqref{tc}, the turning point is located at the point with ${\rm Re}\:t \gg 1$, which would mean, the creation of the particle takes place at late time. For sufficiently large $k$, the production time on the real axis is large enough to regard $\langle\phi^2\rangle\sim v^2$, and the frequency is effectively given by 
\begin{equation}
    \omega_k(t) \sim \sqrt{k^2e^{-2Ht}+ M^2},
\end{equation}
and therefore, the Bogoliubov coefficient would become that in \eqref{betak}.

Let us consider the production of low frequency mode. We approximate the frequency $\omega_k$ as follows:
\begin{align}
    \omega_k(t) &= \sqrt{k^2e^{-2Ht}+\frac12 M^2\left(1+\tanh\left(\frac12\Lambda t\right)\right)}\nonumber\\
&=e^{\frac{\Lambda t}{2}}\sqrt{k^2e^{-2Ht-\Lambda t}+ \frac{M^2}{1+e^{\Lambda t}}}\nonumber\\
&\sim e^{\frac{\Lambda t}{2}}\sqrt{k^2e^{-2Ht}+ M^2}.
\end{align}
Here, we have used the approximations $e^{-2H t-\Lambda t} \sim e^{-2H t}$ and $\frac{1}{1+e^{\Lambda t}} \sim 1$. The former is justified since $\Lambda\ll H$, and the latter is consistent as long as $-\Lambda {\rm Re}\:t \gg 1$.\footnote{Here, we have assumed $e^{\Lambda t} \sim 0$, and this cannot be justified if we are interested in ${\rm Re}t>0$. However, as long as we are interested in the time scale shorter than $\Lambda^{-1}$($\gg H^{-1}$), we can approximate $e^{\Lambda t}\sim C= const.$ and in such a case, we can use the approximated formula by replacing $M^2$ with $M^2(1+C)^{-1}$.}
With this approximation, the turning point is the same as that in \eqref{tc}. 

With the approximated frequency, we find the singulant to be
\begin{align}
    F_k(t)=&{\rm i}\int^t_{t_c}\omega_k(t')dt'\nonumber\\
=&\left[\frac{2{\rm i} e^{\frac{\Lambda  t}{2}} \sqrt{k^2 e^{-2 H t}+M^2} \left(M^2 e^{2 H t}+k^2\right) \,
   _2F_1\left(1,\frac{\Lambda }{4 H}+1;\frac{2 H+\Lambda }{4 H};-\frac{e^{2 H t} M^2}{k^2}\right)}{k^2 (\Lambda -2
   H)}\right]_{t_c}^t
\end{align}
The most relevant quantity is $F_k(t_c^*)$ given by
\begin{equation}
    F_k(t_c^*) \sim -\frac{ \pi  M}{H}\left( 1+\frac{\Lambda   \left(\log \left(\frac{k}{M}\right)+1\right)}{2 H}\right),
\end{equation}
where we have taken the leading order term in $\Lambda$.
Therefore, the Bogoliubov coefficient is asymptotically given by
\begin{equation}
    \beta_k\sim {\rm i}e^{F_k(t_c^*)}
={\rm i}e^{-\frac{\pi M}{H}\left(1+\frac{\Lambda}{2H}\right)}\left(\frac{M}{k}\right)^{\frac{\pi M \Lambda}{2H^2}},
\end{equation}
and the particle number density is given by
\begin{equation}
    n_k \sim e^{-\frac{2\pi M}{H}\left(1+\frac{\Lambda}{2H}\right)}\left(\frac{M}{k}\right)^{\frac{\pi M \Lambda}{H^2}}.
\end{equation}
Thus we have found the produced particle spectrum corrected by the tunneling dynamics. There is an extra factor depending on momentum, which becomes larger for smaller momentum $k$. This seems reasonable since the low frequency modes are more sensitive to the change of the effective mass, particularly because $\chi$ is originally a massless particle. However, we should also note that we have used the coarse-grained expression for $\langle\phi(t)\rangle$, and for very small momentum modes, the wavelength can be larger than the coarse-graining scale roughly given by $H^{-1}$. In such a case, the stochastic inflation might not be an appropriate formalism to describe the production of $\chi$. In this sense, the divergent behavior of $k\to0$ would be corrected by more appropriate formalism. We also note that if we are interested in total number density of the produced particle, the IR divergence would not spoil because the negative power of $k$ given by the tunneling dynamics correction is so small that the phase space volume $\int d^3k$ would cancel the negative power of $k$, and there would not be any IR divergence.

Finally, let us comment on the particle production in the de Sitter spacetime. In the de Sitter space, the notion of particle or its production would not be useful because the frequency never becomes stationary. In a realistic scenario, inflationary de Sitter phase will end at some point and the Universe eventually asymptotes to the present Universe, where the particles would have (almost) time-independent frequency. In such a case, the production of particles would be physically meaningful. Even in such a case, our discussion would not be altered much, and there would be corresponding turning points at which particle production takes place. There, our estimation of the particle production due to tunneling would be useful.

\section{Summary and discussion} \label{sec:coda}
We have analyzed particle production with a tunneling background field by the Stokes phenomenon method and the real time formalism in the flat and in the de Sitter spacetime. 

In the case of the flat spacetime, analyzing the Stokes line, we have found that particle production is efficient only when the tunneling background field is around each vacuum. We have obtained the number density of produced particle \eqref{result_nchi}. This result is a novel consequence of the real time formalism of tunneling \cite{Blanco-Pillado:2019xny} and valid not only for a double-well potential but also for a broad class of potential with at least two metastable vacua. We should note that we have used homogeneous approximation for the tunneling scalar, which causes the continuous oscillation of the tunneling scalar between true and false vacuum. We expect that taking the spatial dependence would change our result slightly, but not so significantly because we have considered a heavy scalar whose Compton wavelength is much shorter than the nucleated bubble radius.

In the case of the de Sitter spacetime, since it is known that particle production occurs without any background dynamics, first we have revisited such particle production by using the Stokes phenomenon method. Although the total produced particle number density after comoving momentum integration suffers from divergence similarly to the previous studies, we have shown that a physical momentum distribution is convergent. Comparing with this result, we have also analyzed particle production with a tunneling background field using the stochastic inflation formalism, which gives a time-dependent distribution of a scalar field value in the de Sitter spacetime. We have found that the efficiency of production for low momentum modes is enhanced by power-law due to the background tunneling dynamics. The resultant particle number density also seems to suffer from the IR divergence; however, it results from the pure (eternal) de Sitter phase and thus it is not problematic in realistic inflationary models where inflationary phase will end at some point.

We have found different spectra of produced particle in these cases as we may expect from the difference of geometry. We should note that the vacuum decay processes in flat and de Sitter spacetime are rather different. In flat spacetime case, the vacuum decay is caused by some fraction of large initial velocity component, and therefore, the dynamics of tunneling field is relatively fast, once the tunneling region has acquired a large enough momentum to cross over the barrier (which may require a long time to realize if the transition rate is small). Therefore, up to overall factor, the produced particle spectrum is almost the same as a scalar field with an oscillating mass. On the other hand, in the de Sitter case, the vacuum decay can be understood as the ``thermal" fluctuation or random walk. Therefore, the tunneling field moves to the true vacuum very slowly, where the transition time scale is inverse of the tunneling rate. Such slowness implies that the expansion effect is more responsible for the particle production, and as we have seen in Sec.~\ref{sec:dS}, the produced particle spectrum is similar to the case of a massive free scalar without coupling to a tunneling field. We also note that we are not able to take $H\to 0$ limit, since in such a limit, the stochastic inflation description cannot be applied.

Finally, we have to add one more comment.  In order to apply for Higgs instability one needs to incorporate  gravity because the spacetime would become anti-de Sitter with large negative curvature after vacuum decay. We expect that our discussion with Stokes phenomenon analyses in real time formalism would be applicable to such cases. We intend to consider such extensions in future work.

 \vskip 1cm
\noindent
{\large\bf Acknowledgements}\\
We would like to thank Minxi He, Kohei Kamada, Takumi Hayashi, and Hiroaki Tahara for useful discussions and comments. SH is supported by JSPS KAKENHI, Grant-in-Aid for JSPS Fellows 20J10176 and the Advanced Leading Graduate Course for Photon Science (ALPS). YY is supported by JSPS KAKENHI, Grant-in-Aid for JSPS Fellows JP19J00494. JY is supported by JSPS KAKENHI Grant Nos.~15H02082, 20H00151 and Grant-in-Aid for Scientific Research on Innovative Areas 20H05248.

\appendix
\section{Intersection points between the Stokes lines and the real time axis} \label{app}
\begin{figure}[htbp]
	\centering
	\includegraphics[width=.64\textwidth]{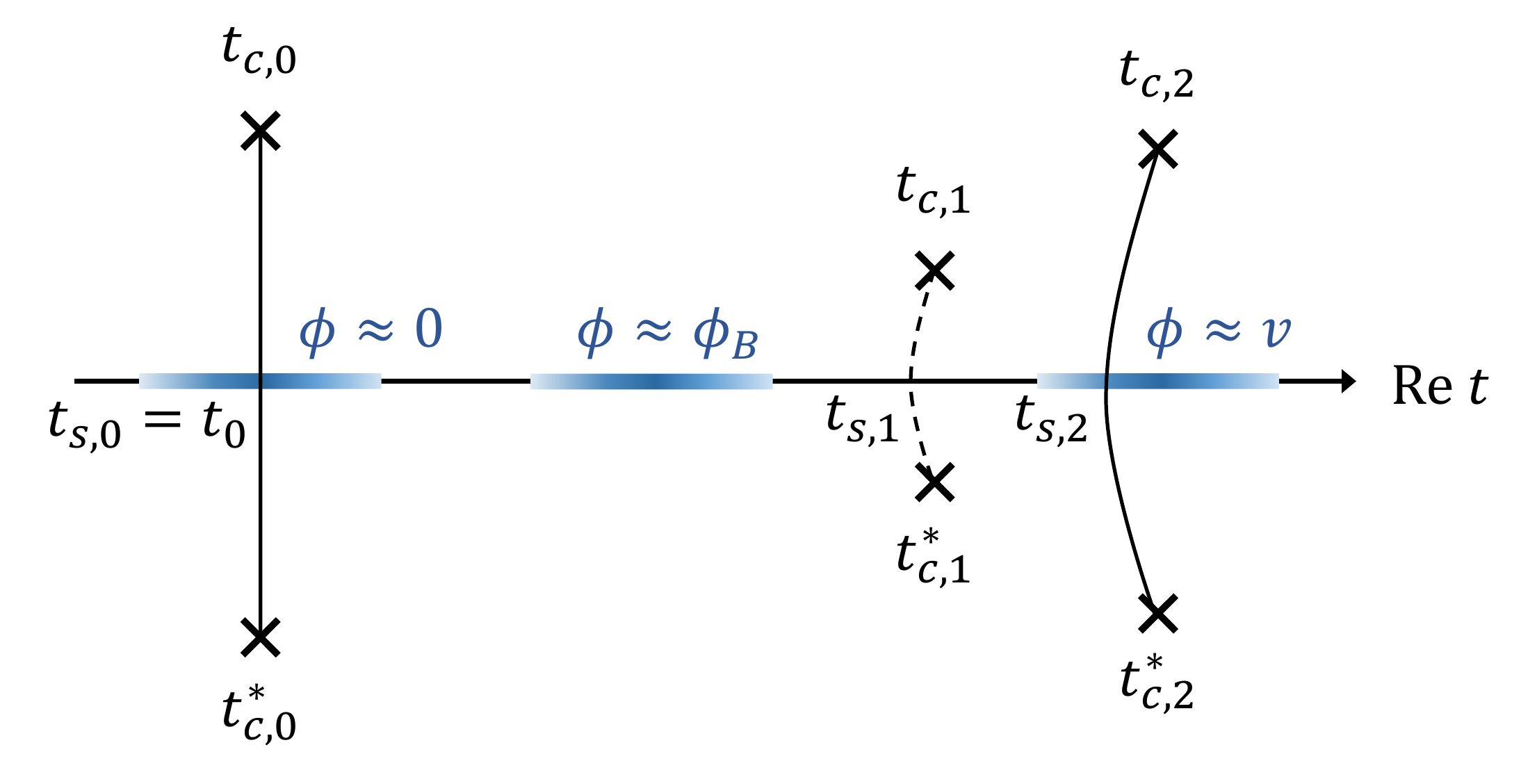}
	\caption{\label{fig:SL} A schematic picture of the Stokes lines in our case. Cross marks and blue shaded regions denote each turning point and associated region where the approximations \eqref{phiapproxA} are valid. The Stokes line depicted as a dashed line is irrelevant for particle production since it is out of the range of the  approximation. The relevant Stokes lines intersect the real time axis only at the points where $\phi$ is around each vacuum.}
\end{figure}

In this appendix, we discuss the structure of the Stokes lines in complex time plane in order to know which Stokes line is relevant for particle production in Sec.~\ref{sec:flat}. A schematic picture of these Stokes lines is shown in Fig.~\ref{fig:SL}, which indicates that $\chi$-production occurs only when $\phi$ is around each vacuum. Below we will discuss the details of the Stokes lines.

As we have discussed in Sec.~\ref{sec:flat}, $V(\phi)$ is expanded around each vacuum and around the barrier as
\begin{equation}
    V(\phi) = \begin{cases}
        \varepsilon + \frac{1}{2} m_F^2 \phi^2 + \mathcal{O}(\phi^3) & (\phi \approx 0) \\
        h - \frac{1}{2} m_B^2 (\phi - \phi_B)^2 + \mathcal{O}((\phi - \phi_B)^3) & (\phi \approx \phi_B) \\
        \frac{1}{2} m_T^2 (\phi - v)^2 + \mathcal{O}((\phi - v)^3) & (\phi \approx v)
    \end{cases},
\end{equation}
where $m_B^2 = V''(\phi_B), m_T^2 = V''(v)$, the motion of $\phi$ inside $S_B$ is approximated, respectively, as
\begin{equation}
    \phi(t) \approx \begin{cases}
        \frac{\dot{\phi}_0}{m_F} \sin\left[ m_F (t - t_0) \right] & (\phi \approx 0) \\
        \frac{\sqrt{\dot{\phi}_0^2 - 2(h - \varepsilon)}}{m_B} \sinh\left[ m_B (t - t_1) \right] + \phi_B & (\phi \approx \phi_B) \\
        \frac{\sqrt{\dot{\phi}_0^2 + 2\varepsilon}}{m_T} \sin\left[ m_T (t - t_2) \right]+v & (\phi \approx v)
    \end{cases}. \label{phiapproxA}
\end{equation}
Here we neglect the back-reaction from $\chi$-particle production, which is justified by the smallness of the production rate. We also consider the dynamics around the barrier for verifying that $\chi$-production does not occur there.

For discussion in Sec.~\ref{sec:flat}, we have to check if the Stokes lines derived in each region, namely $\phi \sim 0,\phi_B,v$, actually intersect the real time axis within this range where the each approximation is valid.
In order to find $t_s$ on the Stokes line satisfying ${\rm Im}F_k(t)=0$, we decompose the integration contour of the singulant \eqref{singulant} into the contour vertical to and along the real time axis as\footnote{As long as there is no pole or branch cut between two turning points, we may deform the integration contour.}
\begin{equation}
    \mathrm{Im}\:F_k(t_s) = 2\underbrace{\mathrm{Re}\int_{t_c}^{\mathrm{Re}\:t_c} dt\: \omega_k(t)}_{\equiv \Delta} + 2\mathrm{Re}\:\int_{\mathrm{Re}\:t_c}^{t_s} dt\: \omega_k(t) = 0,
\end{equation}
and therefore $t_s$ is generally approximated as
\begin{equation}
    t_s \approx \mathrm{Re}\:t_c - \frac{\Delta}{\omega_k({\mathrm{Re}\:t_c})}.
\end{equation}
In the region around the false vacuum $\phi\approx 0$,
\begin{align}
    \Delta_0 &= \int_0^{\mathrm{Im}\:t_{c,0}} \mathrm{Im}\left[ \omega_k(\mathrm{Re}\:t_{c,0} + {\rm i}\tau) \right] d\tau \nonumber \\
    &\approx \int_0^{\sinh^{-1}x} \mathrm{Im} \sqrt{\frac{g\dot\phi_0^2}{m_F^2} (x^2 - \sinh^2 \xi)} m_F^{-1} d\xi = 0, \quad (\xi = m_F \tau)
\end{align}
and hence, the Stokes line is the straight line connecting $t_{c,0}$ and $t_{c,0}^*$, from which $t_{s,0}=t_0$ follows. Substituting the expression of $\phi$ around the barrier $\phi \approx \phi_B$~\eqref{phiapproxA} into $\Delta=\mathrm{Re}\int_{t_c}^{\mathrm{Re}\:t_c} dt\: \omega_k(t)$, we find
\begin{align}
    \Delta_1 &= \int_0^{\mathrm{Im}\:t_{c,1}} \mathrm{Im}\left[ \omega_k(\mathrm{Re}\:t_{c,1} + {\rm i}\tau) \right] d\tau \nonumber \\
    &\approx \int_0^{\frac\pi2} \frac{\sqrt{k^2 + M_0^2}}{m_B} \mathrm{Im}\sqrt{1 + \left( \sqrt{1 - \frac{1}{y^2}} \cos\xi + {\rm i} \sin\xi + \sqrt\frac{g\phi_B^2}{k^2 + M_0^2} \right)^2} d\xi \nonumber \\
    &\approx \frac{\sqrt{k^2 + M_0^2}}{m_B}\left( \int_0^{\frac\pi2}  \mathrm{Im}\sqrt{1 + e^{2{\rm i}\xi}} d\xi\right)\label{delta1}
\end{align}
at the zeroth order in $y^{-1}$ and $\sqrt\frac{g\phi_B^2}{k^2 + M_0^2}$. After performing numerical integration in the parentheses in the last equality of \eqref{delta1}, we obtain
\begin{align}
    &\Delta_1 \approx 0.53 \times \frac{\sqrt{k^2 + M_0^2}}{m_B} \\
    &\Rightarrow t_{s,1} \approx t_1 + [\ln(2y) - 0.53] m_B^{-1}.
\end{align}
However, this $t_{s,1}$ is out of the range $\phi \approx \phi_B \Leftrightarrow m_B(t - t_1) < 1$ since $y \gg 1$. Therefore, this Stokes line should be regarded as unphysical one, and we neglect it. This shows that particle production indeed does not take place near the barrier. We substitute $\phi$ around the true vacuum $\phi \approx v$ in \eqref{phiapproxA} into the expression of $\Delta$, which yields
\begin{equation}
    \Delta_2 \approx \sqrt{k^2 + M_0^2} \int_0^{\sinh^{-1}z} \mathrm{Im} \sqrt{1 - \frac{\sinh^2 \tau}{z^2} + {\rm i}\frac2z \delta \left( 1 - \frac{\cosh \xi}{z\sqrt{1 + z^2}} \right)} m_T^{-1} d\xi.\label{delta2}
\end{equation}
Again, we numerically look for an approximated expression of \eqref{delta2} for various $z$ and $\delta < 1$ and find
\begin{equation}
    \Delta_2 \approx \frac{\sqrt{k^2 + M_0^2}}{m_T} \mathcal{C},
\end{equation}
where
\begin{equation}
    \mathcal{C} \begin{cases}
        \lesssim 0.1 & (z<1) \\
        \approx \delta & (z>1)
    \end{cases}.
\end{equation}
Hence, the Stokes line intersects the real time axis at
\begin{equation}
    t_{s,2} \approx t_2 - \left( \frac{\mathcal{C}}{\sqrt{1 + \delta^2}} + \frac{\delta}{\sqrt{1 + z^2}} \right) m_T^{-1},
\end{equation}
which is actually within the range $\phi \approx v \Leftrightarrow m_T(t - t_2) < 1$ since we have assumed $\delta \ll 1$. If $\delta$ is much larger than unity, one finds that $t_{s,2}$ is also out of the range of approximation and this Stokes line is irrelevant. This seems physically reasonable since, for $\delta \gg 1$, $\chi$ becomes much heavier after the vacuum decay and the production rate of such a heavy particle should be suppressed. 

\section{Time dependent expectation value from probability distribution function in de Sitter spacetime}
\begin{figure}[bp]
\centering
\includegraphics[width=.72\textwidth]{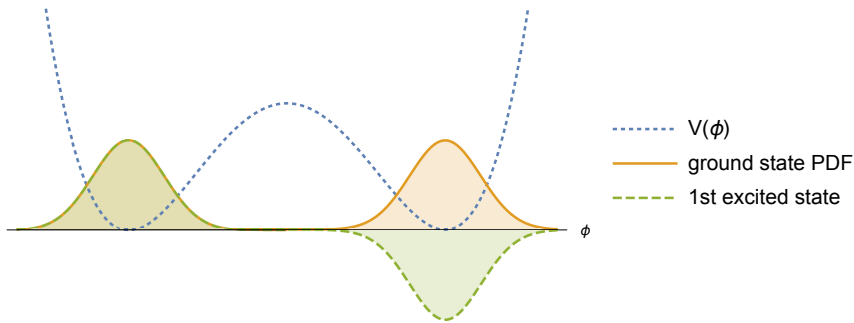}
\caption{\label{fig:double_well_1} Eigenstates of $\phi$ which has a well-separated double-well potential.}
\end{figure}

\begin{figure}[htbp]
\centering
\includegraphics[width=.72\textwidth]{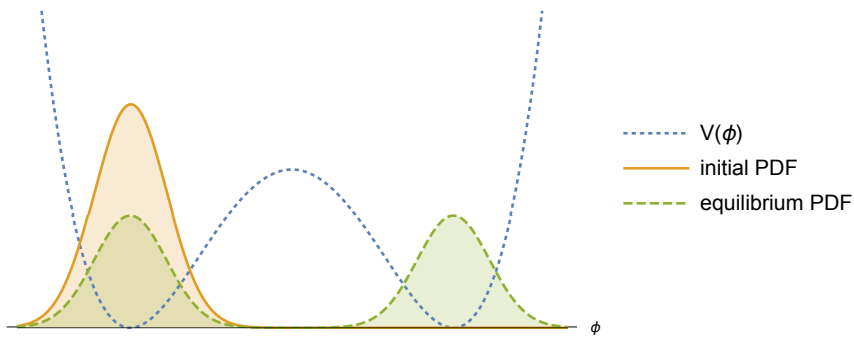}
\caption{\label{fig:double_well_2} The initial and final probability distribution functions. In this case, the final distribution is not localized at one minimum since two minima are degenerated.}
\end{figure}

Here we briefly discuss how the expectation value in \eqref{vev} is derived from the probability distribution function \eqref{pdf} with referring to~\cite{Starobinsky:1994bd}. Since we consider the case where the energy difference between false and true vacua is sufficiently small, we may approximate the situation as follows: We assume the scalar field $\phi$ has a double-well potential
\begin{align}
	V(\phi) = \frac{\lambda}{4} \left[ \left( \phi - \phi_{\rm off} \right)^2 -  \phi_0^2 \right]^2 \label{potential}
\end{align}
in de Sitter space. If two minima $\phi = \phi_{\rm off} \pm \phi_0$ are sufficiently separated ($\sqrt{\lambda}\phi_0^2 \gg H^2$), eigenstates of $\phi$ can be approximated as superposition of those of a massive non-interacting scalar field around each minimum (Fig.~\ref{fig:double_well_1}).
Therefore, the eigenfunctions of the ground state and the first excited state can be written as
\begin{align}
	\Phi_0 (\phi) &= \frac{1}{\sqrt{2}} [\Phi_\ast (\phi - (\phi_{\rm off} - \phi_0)) + \Phi_\ast (\phi - (\phi_{\rm off} + \phi_0))], \\
	\Phi_1 (\phi) &= \frac{1}{\sqrt{2}}  [\Phi_\ast (\phi - (\phi_{\rm off} - \phi_0)) - \Phi_\ast (\phi - (\phi_{\rm off} + \phi_0))],
\end{align}
respectively. Here $\Phi_\ast (\phi)$ is the ground state around each minimum:
\begin{equation}
	\Phi_\ast (\phi) = \frac{1}{\sqrt{N}} \exp\left( - \frac{4 \pi^2}{3 H^4} V_0(\phi) \right),
\end{equation}
where $V_0(\phi) = \lambda \phi_0^2 \phi^2$ and $N$ is a normalization factor. Let us take the initial condition as that $\phi$ is localized around the minimum $\phi = 0$ at $t=\tau$ (Fig.~\ref{fig:double_well_2}), and then the probability distribution function~\eqref{pdf} becomes
\begin{equation}
	\rho(\phi, t) = \exp\left( - \frac{4 \pi^2 V(\phi)}{3 H^4} \right) \times \frac{1}{\sqrt{2}} \left[ \Phi_0 (\phi) + \Phi_1 (\phi) e^{-\Lambda_1 (t-\tau)} \right]. \label{1PDF}
\end{equation}
Using this one-point probability distribution function, the expectation value of $\phi^2$ can be calculated as
\begin{align}
	\Braket{\phi^2 (t)} &= \int_{-\infty}^\infty d\varphi \; \varphi^2 \rho(\varphi, t) \nonumber \\
	&= \left\{ \frac{1 + e^{-\Lambda_1 (t-\tau)}}{2} \int_{-\infty}^\infty d\varphi \; \varphi^2 \frac{1}{\sqrt{N}} \exp\left( - \frac{4 \pi^2}{3 H^4} \left[ V(\varphi) + V_0(\varphi - (\phi_{\rm off} - \phi_0)) \right] \right) \right. \nonumber \\
	&\quad \left. + \frac{1 - e^{-\Lambda_1 (t-\tau)}}{2} \int_{-\infty}^\infty d\varphi \; \varphi^2 \frac{1}{\sqrt{N}} \exp\left( - \frac{4 \pi^2}{3 H^4} \left[ V(\varphi) + V_0(\varphi - (\phi_{\rm off} + \phi_0)) \right] \right) \right\}.
\end{align}
Since the two minima are well-separated, we can approximate the integrands by $V(\varphi) + V_0(\varphi - (\phi_{\rm off} \pm \phi_0)) \approx 2V_0(\varphi - (\phi_{\rm off} \pm \phi_0))$ and obtain;
\begin{align}
	\Braket{\phi^2 (t)} &\approx \frac{1 + e^{-\Lambda_1 (t-\tau)}}{2} (\phi_{\rm off} - \phi_0)^2 + \frac{1 - e^{-\Lambda_1 (t-\tau)}}{2} (\phi_{\rm off} + \phi_0)^2 \nonumber \\
	&= (\phi_{\rm off}^2 + \phi_0^2) - 2\phi_{\rm off} \phi_0 e^{-\Lambda_1 (t-\tau)}.
\end{align}
Substituting $\phi_{\rm off} = \phi_0 = v/\sqrt2$ into this equation, we obtain \eqref{vev}. Note that the final (equilibrium) state is the ground state, and hence $\phi$ is {\it not} localized at one minimum but equally distributed to two minima at $t\to\infty$ as shown in Fig.~\ref{fig:double_well_2}. However, this superposition originates from our double well approximation, and if the energy difference between the false and the true vacuum is taken into account, we expect the ground state would be localized at true one.

\bibliographystyle{JHEP}
\bibliography{ref}
\end{document}